\documentclass{pic2012}
\usepackage{mathtools}
\def\runauthor{Kai Zhu}
\def\shorttitle{charmonium and light hadron spectrum}
\begin{document}

\title{Charmonium and Light Meson Spectroscopy}
\author{PIC~2012~Kai~Zhu}
\address{Institute of High Energy Physics, CAS\\
Beijing 100049, China\\
E-mail: zhuk@ihep.ac.cn}

\maketitle

\abstracts{This talk reviews recent experimental results on
selected topics in the spectroscopy of charmonia,
charmonium-like states and light mesons.}

\section{Introduction}
In modern high energy physics it has been accepted
generally that quarks are basic building blocks of matter
and Quantum Chromodynamics (QCD) describes their
strong interactions that lead to bound states.
States composed of only light quarks, i.e. up, down
or strange quarks, include light mesons and baryons, such
as $\pi$, $K$, $\rho$, $\eta$, $\eta'$, $\Lambda$, $p$ or
$n$ etc. Using heavy quark-anti-quark pairs, one makes
so-called charmonium or bottomonium, composed of
charm and anti-charm or beauty and anti-beauty quark pairs,
respectively.
Finally, heavy-light quark pairs will form mesons
such as $D$, $B$, $D_s$ and $B_c$ with the charm-up,
charm-down, beauty-up, beauty-down, charm-strange,
and beauty-charm quark combinations.

However, this simple prospect cannot explain the whole
phenomenology of states that has been observed.
In practice, while
it works well in the perturbative, i.e. very high-energy,
region, QCD can not easily describe the non-perturbative strong
interactions when the energy is relatively low.
Physicists often turn to various phenomenological models,
such as quark models, potential models (most of the
potentials to describe charmonium are
modifications of the famous Cornell potential
$V(r)=-k/r+r/a^2$), QCD sum rules, and lattice QCD, etc.
But none of them is really satisfactory yet,
indicating that we don't understand the strong interaction
well when the energy is low.  Thus,
we seek further experimental input to provide more information to
help understand the interaction mechanism of the
non-perturbative strong interaction.

Experiments at the $\tau$-charm energy region are ideal to study
the non-perturbative strong interactions due to their energy
scale. At present there are so-called charm factories such
as the CLEO-c and BES-III experiments working at this energy
region. Other experiments for example $B$ factories such as
BaBar and Belle, the $\phi$ factory KLOE, and $p\bar{p}$ ($pp$)
collider experiments such as CDF-II, D0, LHCb and CMS, can help
with data from other production processes.
This talk will review the results mainly from these
experiments mentioned above; I apologize for not covering everything.

\section{Charmonia and charmonium-like states}
\begin{figure}[!hbt]
\begin{center}
\includegraphics[width=0.9\textwidth]{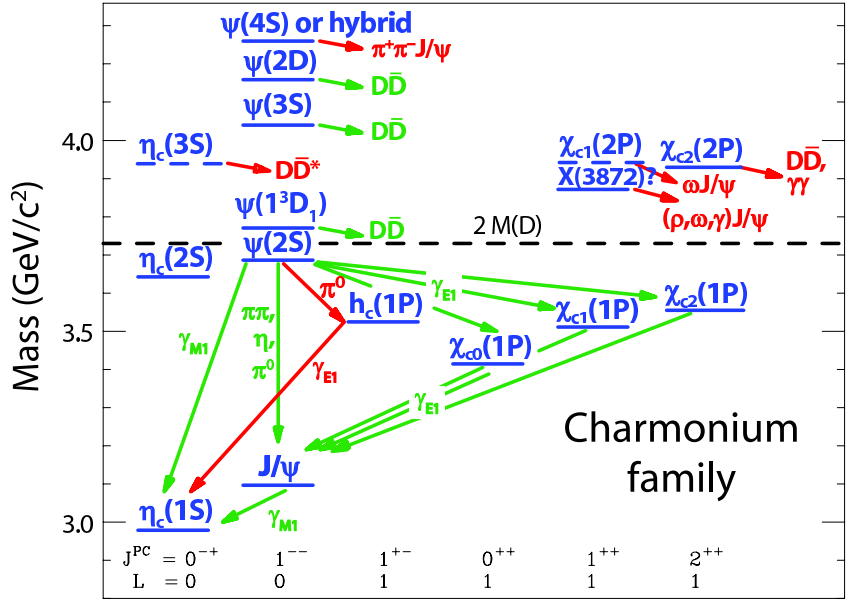}
\caption{\label{fig:charmonium}
Known charmonium states and candidates, with
selected decay modes and transitions. The dashed line is
the open-charm threshold.}
\end{center}
\end{figure}

Charmonium states are made of a charm and an anti-charm quark.
Figure~\ref{fig:charmonium} is
from Ref.~\cite{Eichten:2007qx} and shows the present
status of charmonium $(c \bar c)$ levels. Also
Table~\ref{tab:charmonium} lists known charmonium states
from PDG2012~\cite{pdg2012}.

\begin{table}[!htb]
\begin{tabular}{ccccccc}
\hline \hline
$n$   &   $L$  & $J^{PC}$  &  $n^{2S+1}L_J$ & Name         &  Mass(MeV)  &  Width(MeV) \\
\hline
1 & 0 & $0^{-+}$ & $1^1S_0$ & $\eta_c(1S)$ &  $2981.0\pm1.1$& $29.7\pm1.0$\\
1 & 0 & $1^{--}$ & $1^3S_1$ & $J/\psi     $ &  $3096.916\pm0.011$& $92.9\pm2.8$keV\\
1 & 1 & $0^{++}$ & $1^3P_0$ & $\chi_{c0}(1P)$ &  $3414.75\pm0.31$& $10.4\pm0.6$\\
1 & 1 & $1^{++}$ & $1^3P_1$ & $\chi_{c1}(1P)$ &  $3510.66\pm0.07$& $0.86\pm0.05$\\
1 & 1 & $2^{++}$ & $1^3P_2$ & $\chi_{c2}(1P)$ &  $3556.20\pm0.09$& $1.98\pm0.11$\\
1 & 1 & $1^{+-}$ & $1^1P_1$ & $h_c(1P)$ &  $3525.41\pm0.16$& $<1$\\
1 & 2 & $1^{--}$ & $1^3D_1$ & $\psi(3770)$ &  $3773.15\pm0.33$& $27.2\pm1.0$\\
2 & 0 & $0^{-+}$ & $2^1S_0$ & $\eta_c(2S)$ &  $3638.9\pm1.3$& $10\pm4$\\
2 & 0 & $1^{--}$ & $2^3S_1$ & $\psi(2S)$ &  $3686.109\substack{+0.012\\-0.014}$& $304\pm9$keV\\

  &   & $?^{?+}$ &          & $X(3872)$ &  $3871.68\pm0.17$& $<1.2$\\
  &   & $?^{?+}$ &          & $X(3915)$ &  $3917.5\pm2.7$& $27\pm10$\\
2 & 1 & $2^{++}$ & $2^3P_2$ & $\chi_{c2}(2P)$ &  $3927.2\pm2.6$& $24\pm6$\\
3 & 0 & $1^{--}$ & $3^3S_1$ & $\psi(4040)$ &  $4039\pm1$& $80\pm10$\\
2 & 2 & $1^{--}$ & $2^3D_1$ & $\psi(4160)$ &  $4153\pm3$& $103\pm8$\\
  &   & $1^{--}$ &          & $X(4260)$    &  $4263 \substack{+8 \\-9}$& $95\pm14$\\
  &   & $1^{--}$ &          & $X(4360)$   &  $4361\pm13$ & $74\pm18$\\
4 & 0 & $1^{--}$ & $4^3S_1$ & $\psi(4415)$ &  $4421\pm4$ & $62\pm20$\\
  &   & $1^{--}$ &          & $X(4660)$ &  $4664\pm12$  & $48\pm15$\\
\hline \hline
\end{tabular}
\caption{\label{tab:charmonium} Known charmonium states
from PDG2012.}
\end{table}

\subsection{Conventional charmonium states}
This first part of this talk will focus on recent progress on three
spin-singlet states $\eta_c(1S)$, $h_c(1P)$ and
$\eta_c(2S)$ as well as $\chi_{c2}(2P)$, but will not discuss
other conventional charmonium states such as $J/\psi$,
$\psi(3686)$, $\psi(3770)$, 1P spin-triple $\chi_{c0},
\chi_{c1}$ and $\chi_{c2}$ which have been
measured relatively well already.
We will not discuss the $\psi(4040)(3S)$, $\psi(4160)(2D)$,
$\psi(4415)(4S)$ and the missing 1D
states because there are still controversies about their natures.
However, there will be discussion of the newer charmonium-like
states in the following subsection~\ref{sec:charmonium-like}.

The lightest charmonium state $\eta_c$ has been known
for many years, however the mass and width of this
resonance continue to have large uncertainties when
compared to those of other charmonium states. Notably, there is an
obvious discrepancy between results from radiative
transition and photon-photon fusion processes.

Recently CLEO-c studied the lineshape from $J/\psi\to\gamma
\eta_c$~\cite{Mitchell:2008aa}, and found a fit after
adding a $E_\gamma^3$ form factor to the Breit-Wigner
lineshape, modified by an additional damping term. This
lineshape appears to be different than that in
$\gamma\gamma$ fusion and $h_c\to\gamma\eta_c$.  In
addition, one may also need to carefully consider
interference in the radiative transition to help solve the
$\eta_c$ mass and width ``puzzle''.

\begin{figure}[!hbt]
\begin{center}
\includegraphics[width=0.9\textwidth]{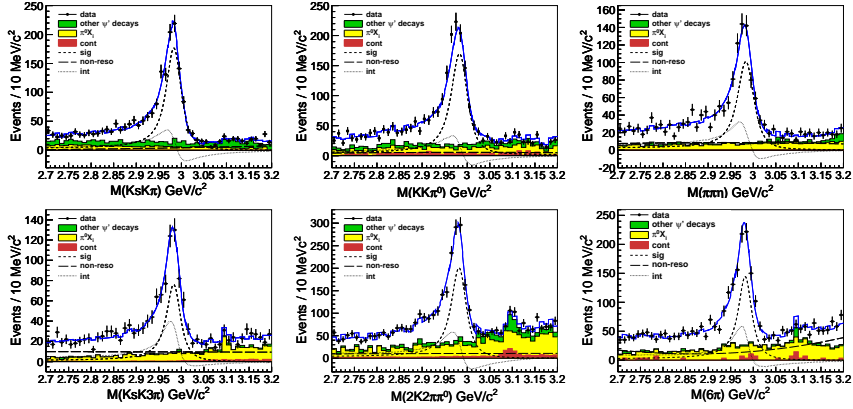}
\caption{\label{fig:combined}Measurement of the mass and width of $\eta_c$ with
considering the interference effect by BESIII.}
\end{center}
\end{figure}

\begin{figure}[!hbt]
\begin{center}
\includegraphics[width=0.45\textwidth]{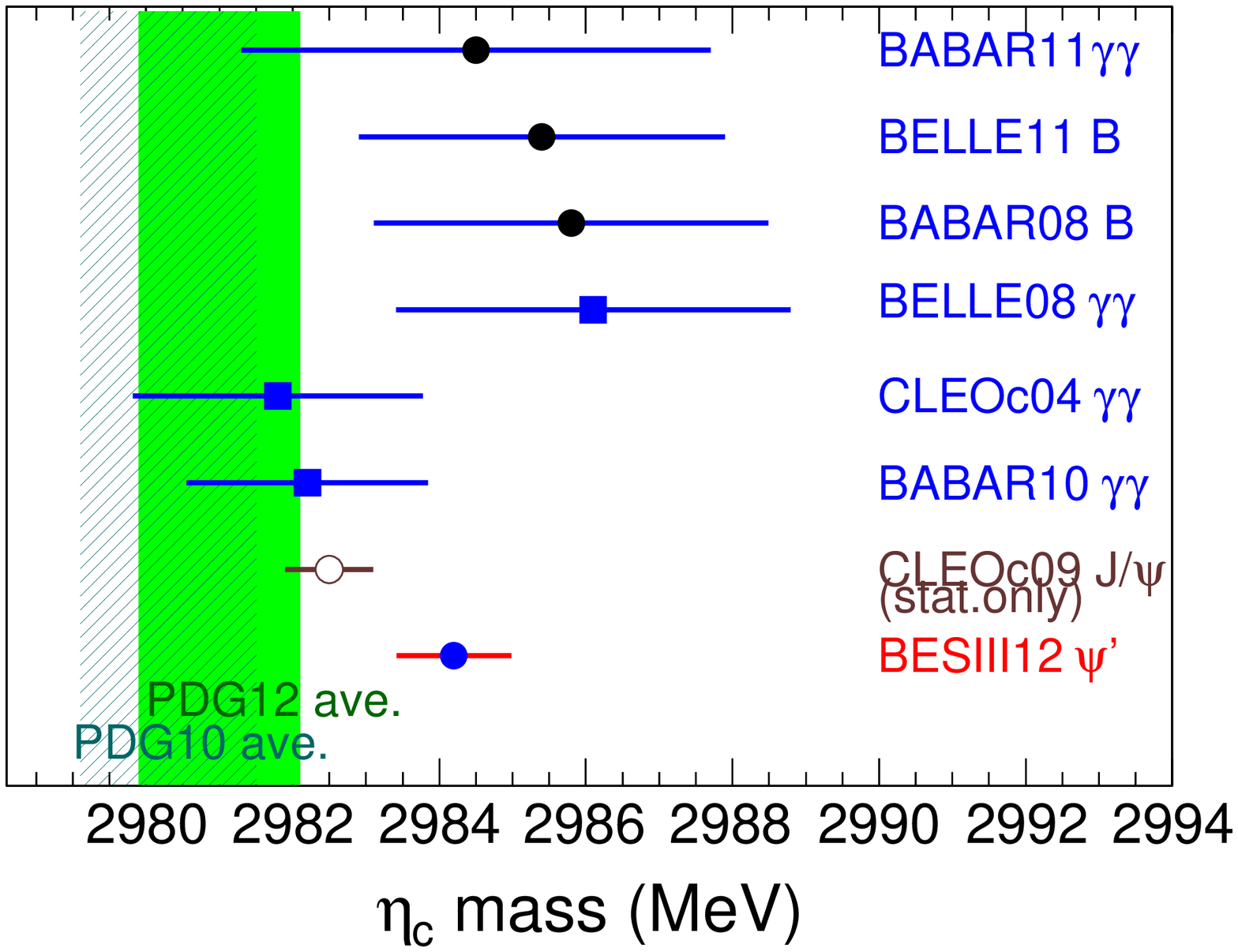}
\includegraphics[width=0.45\textwidth]{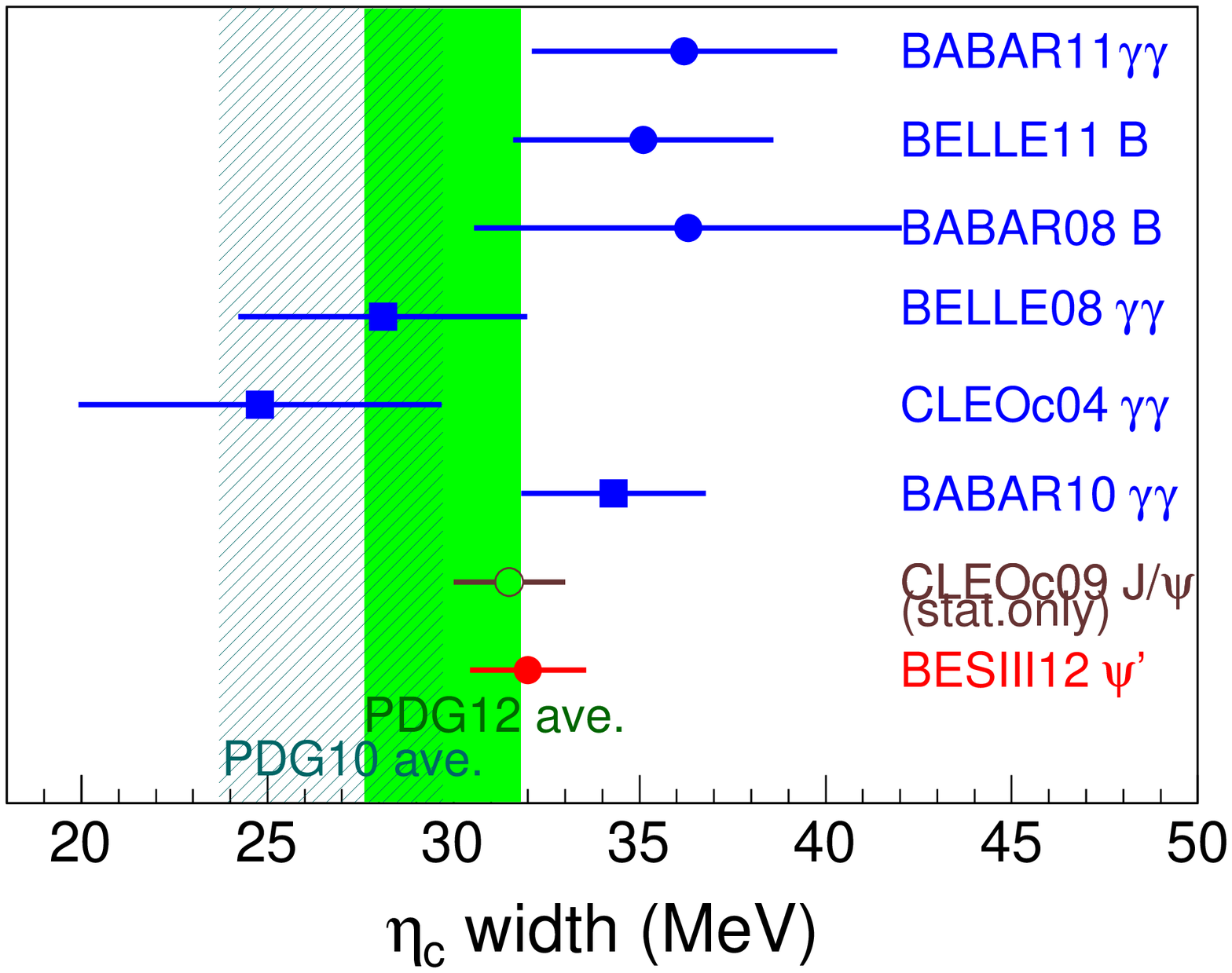}
\caption{\label{fig:etac}Summary of the mass and width of $\eta_c$.}
\end{center}
\end{figure}

Based on this idea, BESIII measured the $\eta_c$ mass and
width, including interference between the amplitudes of resonant
and non-resonant processes~\cite{BESIII:2011ab}; the results are
consistent with the measurement in $\gamma\gamma$ fusion and
calculations from potential models. Figure~\ref{fig:combined}
presents the interference effect in six exclusive decay
modes allowing for a relative phase angle. Belle also
measured $\eta_c$ via $B\to
K\eta_c$~\cite{Vinokurova:2011dy}, in which interference
was considered too as well as a 2D-fit. Other recent
$\eta_c$ measurements are from $\gamma\gamma$ fusion
including $\eta_c \to K_s K \pi$~\cite{Lees:2010de} and $K
K 3\pi$~\cite{delAmoSanchez:2011bt} by BaBar, $\eta_c\to 4\,
\mathrm{prongs}$~\cite{Uehara:2007vb} and
$\eta_c\to\eta'\pi\pi$~\cite{Zhang:2012tj} by Belle.
Figure~\ref{fig:etac} displays a summary of the mass and
width of $\eta_c$ measurements.

BESIII provides the most precise measurement in world at present.
Note that the hyperfine splitting value based on
BESIII results is $\Delta M(1S)=112.5 \pm 0.8$ MeV, which
is consistent with potential model and recent lattice QCD
calculation~\cite{Seth:hadron11}, i.e. $\Delta
M_{hf}(nS)=M(n^3S_1)-M(n^1S_0)=
\frac{32\pi\alpha_s(m_q)}{9}(\phi(0)/m_q)^2 \rightarrow
\Delta M(1S)\approx 118$ MeV.

The $h_c(^1P_1)$ is a singlet 1P wave state, and the first
evidence was from E835 in $p\bar{p}\to h_c \to \gamma
\eta_c$~\cite{Andreotti:2005vu}. Potential models predict
that if there was non-vanishing P-wave spin-spin
interaction, then $\Delta M_{hf}(1P) = M(h_c)- \langle M(1
^3P_J)\rangle \neq 0$. However, from the CLEO-c observation
of $h_c$ in $e^+e^-\to\psi' \to \pi^0 h_c$, $h_c\to \gamma
\eta_c$~\cite{Dobbs:2008ec}, $\Delta M_{hf}(1P) = 0.08 \pm
0.18 \pm 0.12\, \mathrm{MeV/c^2}$, which is consistent with
a very small 1P hyperfine splitting. Further theoretical
predictions can be found in
Refs.~\cite{Kuang:2002hz,Godfrey:2002rp}. Recently, first
measurements of the absolute branching ratios
$\mathcal{B}(\psi' \rightarrow \pi^0 h_c) = (8.4 \pm 1.3
\pm 1.0) \times 10^{-4}$ and $\mathcal{B}(h_c \rightarrow
\gamma \eta_c) = (54.3 \pm 6.7 \pm 5.2)\%$, $M(h_c) =
3525.40 \pm 0.13 \pm 0.18$~MeV/$c^2$ and $\mathcal{B}(\psi'
\rightarrow \pi^0 h_c) \times \mathcal{B}(h_c \rightarrow
\gamma \eta_c) = (4.58 \pm 0.40 \pm 0.50) \times 10^{-4}$
were presented by BESIII~\cite{Ablikim:2010rc}, and
confirmed by CLEO-c with $\mathcal{B}(\psi' \to \pi^0 h_c)
=(9.0\pm1.5\pm1.3)\times 10^{-4}$~\cite{Ge:2011kq}. The
$h_c$ has also been observed in exclusive reactions by
BESIII~\cite{Ablikim:2012ur} and a new production mode $e^+
e^- (4170) \to \pi^+ \pi^- h_c(1P)$ has been found by
CLEO-c~\cite{CLEO:2011aa}.

The $\eta_c(2S)$ was first ``observed'' by Crystal Ball in
1982 via $\psi'\to\gamma X$~\cite{Edwards:1981mq}, however
the mass obtained was much lower than current values.
BESIII has now unambiguously detected this
process~\cite{Ablikim:2012sf}; it is quite experimentally
challenging due to the detection of a soft $50$MeV photon.
It has also been observed in other production mechanisms,
such as via double charmonium
production~\cite{delAmoSanchez:2011bt,Aubert:2005tj}, $B\to
K \eta_c(2S)$~\cite{Choi:2002na}, and $\gamma \gamma \to
\eta_c(2S) \to K K \pi$~\cite{Asner:2003wv,Abe:2007jn}.

\begin{figure}[!htb]
\begin{center}
\includegraphics[width=0.45\textwidth]{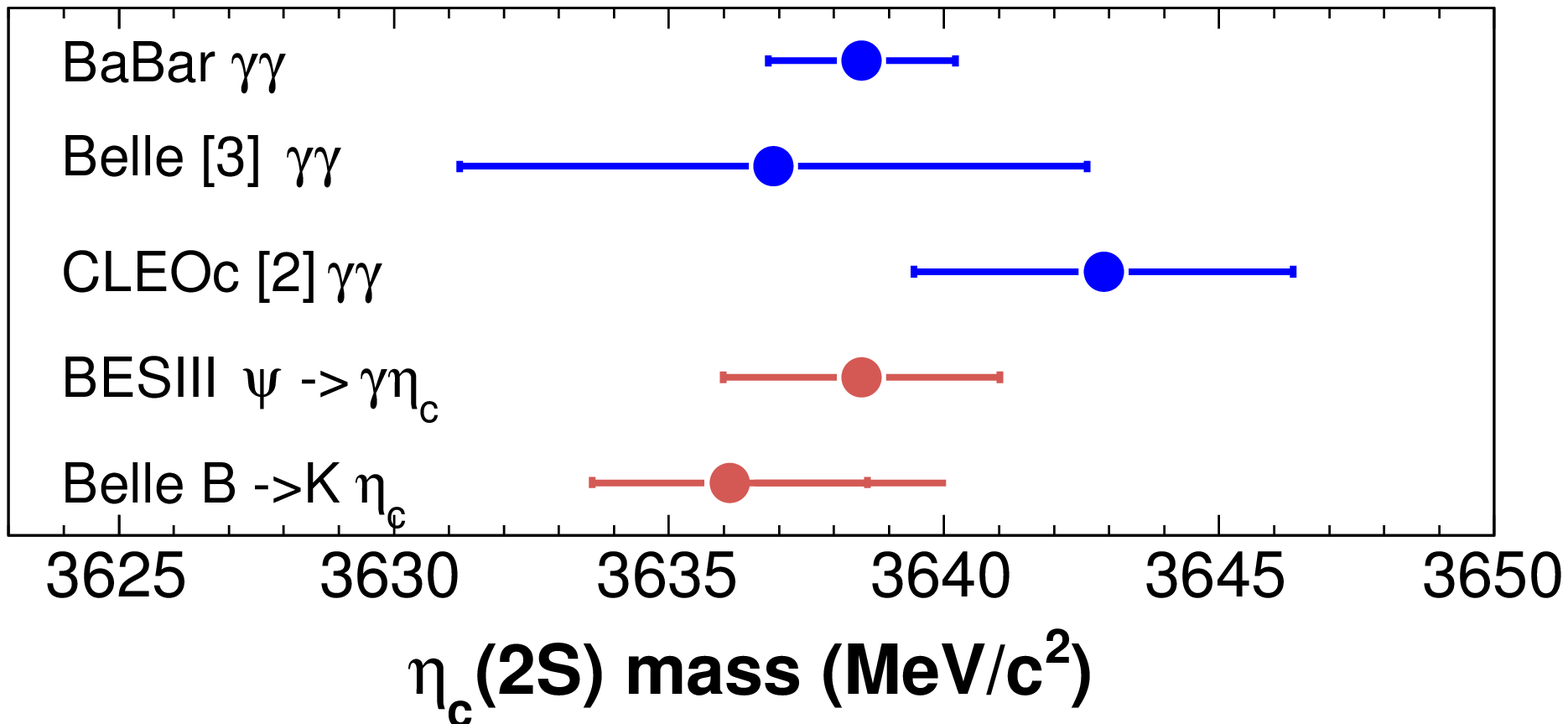}
\includegraphics[width=0.45\textwidth]{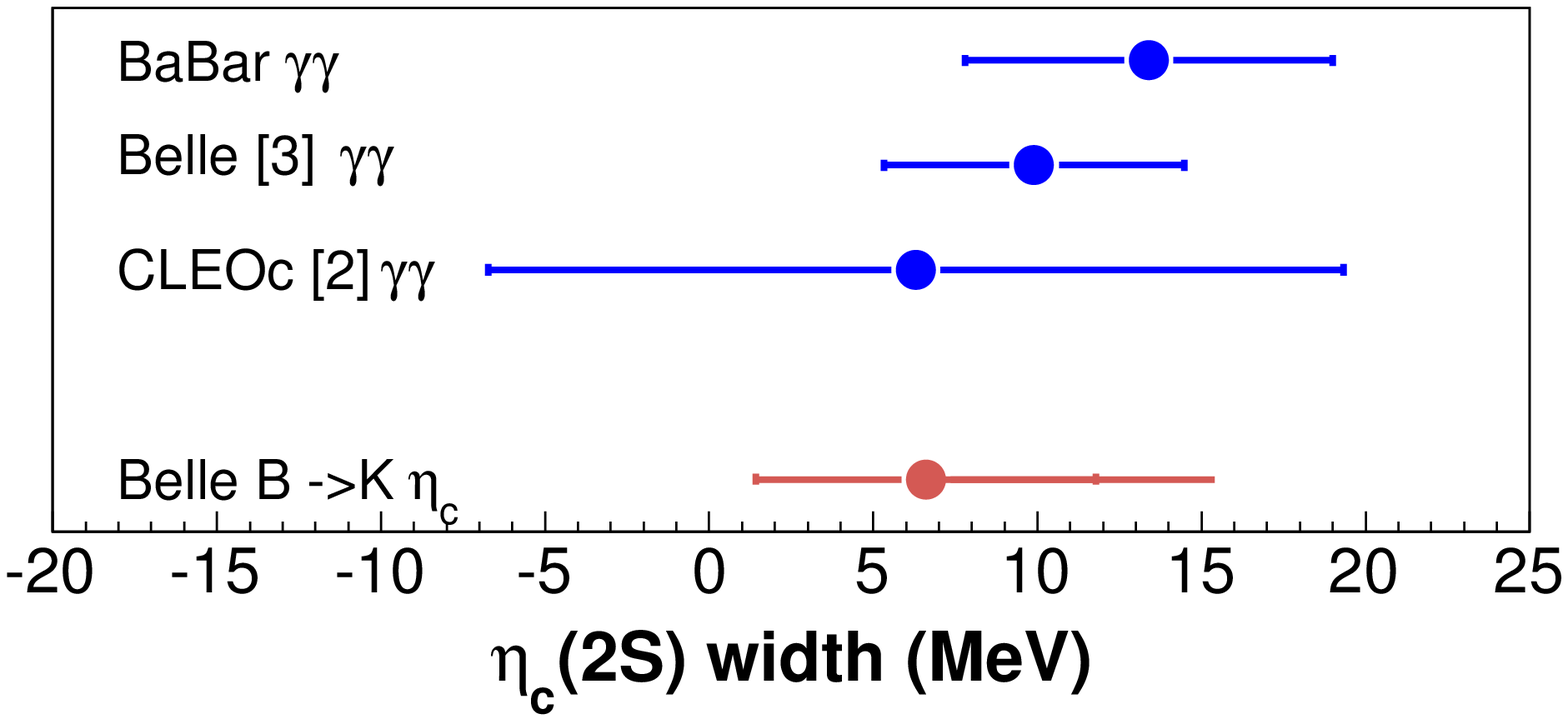}
\caption{\label{fig:etacp}Summary of mass and width of $\eta_c(2S)$.}
\end{center}
\end{figure}

In Ref. \cite{Ablikim:2012sf}, the BESIII collaboration
found the branching fraction of the M1 transition to be
$\mathcal{B}(\psi'\to\gamma\eta_c(2S)) = (6.8\pm
1.1_\mathrm{stat}\pm 4.5_\mathrm{sys})\times 10^{-4}$,
which is consistent with CLEO-c's upper
limit~\cite{CroninHennessy:2009aa} and a potential model
prediction~\cite{Eichten:2002qv}. However, the most precise
mass and width measurement is from $\gamma \gamma$
fusion~\cite{delAmoSanchez:2011bt} by BaBar. There is also
a measurement from $B$ decay~\cite{Vinokurova:2011dy}, in
which it found that interference effects are important.
Figure~\ref{fig:etacp} gives a summary of the mass and
width determinations for the $\eta_c(2S)$.

The $\chi_{c2}(2P)$, previously $Z(3930)$, was first
observed in $B\to K \omega J/\psi$ near the $\omega J/\psi$
threshold~\cite{Abe:2004zs}, and is now also observed in $\gamma
\gamma\to D \bar{D}$ by Belle~\cite{Uehara:2005qd} and
BaBar~\cite{Aubert:2010ab}. At present the
averaged mass and width of $\chi_{c2}(2P)$ from Belle and
BaBar is $M = 3927.2 \pm 2.6$ MeV and $\Gamma = 24 \pm 6$
MeV.

\subsection{Charmonium-like states}
\label{sec:charmonium-like} In this subsection we give
a brief review of the $X(3872)$ (there are already many good reviews
about this resonance) and also summarize recent work on the $X(3823)$
and the newer $1^{--}$ states, including
the $Y(4008)$, $Y(4260)$, $Y(4360)$, $Y(4630)$, $Y(4660)$,
$G(3900)$ along with the discovery of $\psi(4040,4160)\to \eta
J/\psi$. We will not cover the $Y(4160)$ that is
observed in $\phi J/\psi$ by CDF but not confirmed by Belle
and LHCb, or charged $Z$ particles that are not solidly
established yet.

The $X(3872)$ was first observed in $B\to K(J/\psi\pi^+
\pi^-) $ by Belle~\cite{Choi:2003ue}; its mass is very
close to the $D^{*0}\bar{D}^0$ threshold and the width is
less than the experimental resolution.  It was later
confirmed by BaBar, CDF and D0. The quantum numbers of he
$X(3872)$ appear to be $J^{PC} = 1^{++}$ or $2^{-+}$. It
can be produced in $p\bar{p}$ collision or via $B$ decays.
It can decay to open charm final states or to charmonium.
Since its discovery, there have been many theories and
models which try to explain the $X(3872)$, such as
considering it as charmonium state, $D^*\bar{D}$ molecule,
tetra-quark state, or a $c\bar{c}$-gluon hybrid, however
none of them is totally satisfactory and the nature of
$X(3872)$ is still a mystery.

Clear evidence of a signal at $3823\, \mathrm{MeV/c^2}$ via
$B^{\pm} \to X(3823)K^\pm \to \gamma \chi_{c1} K^{\pm}$
with $4.2\sigma$ significance has been reported at
Charm2012~\cite{Bhardwaj:2012ic}, while there is no strong
evidence from $X(3823) \to \gamma \chi_{c2}$. From some
theoretical
predictions~\cite{Eichten:2002qv,Bhardwaj:2012ic,Eichten:2004uh},
this newly observed $X(3823)$ seems to be the missing
$\psi_2$($^3D_2$) state from the charmonium spectrum.

There are many $1^{--}$ states, so called $Y$ states,
observed via the ISR process and located where only one
$c\bar{c}$ vector state should be left in the charmonium
spectrum.  These states are easily seen in decays to $\pi \pi
J/\psi$ or $\pi \pi \psi'$, but there is no sign yet of
$D^{(*)}\bar{D}^{(*)}$ final
states~\cite{Aubert:2005rm,Wang:2007ea,Yuan:2007sj,Pakhlova:2008vn,Pakhlova:2008zza,Abe:2006fj,Pakhlova:2007fq}.
The precise nature of these $Y$ states is still a mystery.

The $G(3900)$
enhancement~\cite{Pakhlova:2008zza,Aubert:2006mi}, another
$1^{--}$ candidate from the ISR process, is located in a mass
region where the quark model does not predict any
corresponding $c\bar{c}$ vector state.  Unlike the
previously mentioned $Y$ states, the $G(3900)$ was observed
decaying into $D\bar{D}$.

A recent analysis at Belle found large, i.e. $\sim 1\%$
level, hadronic transition rates of the $\psi(4040,4160)
\to \eta J/\psi$~\cite{Wang:2012bg}. This is a challenge to
the general belief that these two states are charmonium
states.

\section{Light hadron spectroscopy}
QCD predicts new forms of hadrons in addition to mesons and
baryons, such as multi-quark states and states involving gluons.
These include tetra-quark states ($q\bar{q}q\bar{q}$), penta-quark
states ($qqqq\bar{q}$), hybrids ($q\bar{q}g$, $qqqg$),
and glueballs ($gg$, $ggg$). This section
will cover recent observations of three scalar states
$f_0(1500)$, $f_0(1710)$, $f_0(2100)$ and a tensor
$f'_2(1525)$, as well as $X(1810)$, $X(1835)$ and the first
observation of $\gamma \gamma \to \omega \omega$,
$\phi\phi$, $\omega\phi$. Some of these are exotic
state candidates.

There are preliminary PWA results for $J/\psi\to\gamma
\eta\eta$ at BESIII, in which three scalar states
$f_0(1500)$, $f_0(1710)$, $f_0(2100)$ and a tensor
$f'_2(1525)$ are measured to be the dominant contributions
to the decay rate.  As we know, the $f_0(1500)$ has also
been observed in many different final states such as $\pi
\pi$, $4\pi$, $\eta \eta$, $\eta \eta'$ and $K \bar{K}$
~\cite{Uehara:2008ep,Amsler:2006du,Ablikim:2006db}.
The $f_0(1710)$ is seen in $\pi \pi$, $\omega \omega$, $\eta \eta$
and $K \bar{K}$~\cite{Thomas:2008ga,Uman:2006xb,Ablikim:2006db,Ablikim:2005kp,Ablikim:2004st}
and the $f_0(2100)$ is seen in $\pi \pi$, $\eta \eta$ and
$4\pi$~\cite{Bai:1999mm,Anisovich:2000ut}.

BESII has observed a $M(\omega \phi)$ threshold enhancement
in $J/\psi \to \gamma \omega \phi$~\cite{Ablikim:2006dw}
with $M=1812\substack{+19 \\-26} \pm 18\, \mathrm{MeV/c^2}$
and $\Gamma = 105 \pm 20 \pm 28\, \mathrm{MeV/c^2}$,
with $J^{PC}$ favoring $0^{++}$ over $0^{-+}$ and $2^{++}$.
This $X(1810)$ has been confirmed at BESIII with much
larger statistics in $J/\psi \to \gamma \omega
\phi$~\cite{Ablikim:2012li}.

BESIII also confirmed another discovery from BESII, the
$X(1835)$~\cite{Ablikim:2005um,Ablikim:2010au}, in a sample of
$225$ million $J/\psi$ events.  This analysis examined
$J/\psi\to \gamma \eta' \pi^+ \pi^-$ and also found two new structures
in addition. However, without an amplitude analysis it is very difficult
to interpret these new states.  The $X(1835)$ resonance is confirmed in
the $\gamma \gamma$ fusion process $e^+ e^- \to e^+ e^- \pi \pi
\eta'$~\cite{Zhang:2012tj}. Its mass is determined as
$M=1836.5 \pm 3.0 \substack{+4.7 \\
-2.1}\, \mathrm{MeV/c^2}$~\cite{Ablikim:2010au}, which is very close to the
newly found $p\bar{p}$ enhancement in $J/\psi \to \gamma p
\bar{p}$~\cite{BESIII:2011aa} whose mass is
$M=1832\substack{+19\\-15}(\mathrm{stat.})\substack{+18\\-17}
(\mathrm{sys.})\pm 19 (\mathrm{model})$. However their
widths are very different: it seems the width of $X(1832)$
is around $190$ MeV while for the $p\bar{p}$ enhancement, it is
close to zero. This $p\bar{p}$ enhancement has
been confirmed by CLEO-c in $J/\psi \to \gamma p
\bar{p}$~\cite{Alexander:2010vd}, but has not been observed
in any of $\psi'\to \gamma p
\bar{p}$~\cite{Ablikim:2006pt}, $\Upsilon(1S)\to \gamma p
\bar{p}$~\cite{Athar:2005nu} or $J/\psi\to \omega p
\bar{p}$~\cite{Ablikim:2007ac}.

There is a first measurement of the total cross-section $\sigma_{tot}$ and
also cross-sections for the spin-parity subcomponents $\sigma(J^P = 0^+, 2^+)$
for $\gamma\gamma \to \omega\phi, \phi\phi, \omega\omega$~\cite{Liu:2012eb},
which is a perfect process to study tetra-quark states.  It is found
that the main components of the cross section are scalar
(continuum QCD) and tensor (resonance),
very different from theoretical predictions.

\section{Summary}
Charmonium and light-hadron spectrums provide a platform to
study non-perturbative mechanisms in QCD. In recent years, many
expected and unexpected discoveries were made at charm and
$B$-factories. Especially after BESIII, the experiments have
entered an era in which all the predicted charmonium states
below charm threshold have been well established and the main
work focuses on improvements in precision. However, the
observations of $X, Y, Z$ particles provide us a big challenge
as well as a chance to understand QCD better. At
the same time, many new light resonances have been found,
some from amplitudes analysis. These discoveries raised
natural questions such as are they really new resonances or
are some just the same ones observed before?  What are the parameters:
mass, width and quantum numbers?  What is their
nature: are they glueballs or hybrid states or other new
exotic candidates?

In order to answer these questions, one needs the
contributions from both theorists and experimentalists. For
theory, we may need to update present methods such as
potential models, sum rules, and lattice QCD (will benefit
from super-power computers that may take advantage of the
rapid developing quantum information techniques). Or we may
have to depend on totally novel methods to interpret these
new results. At the same time, fore-front experimental
methods, such as $K$-matrix methods in PWA,
machine-learning techniques, multiple variable analysis
(MVA) and others may be needed to extract more useful
information from present experiments. Furthermore, new and
updated experiments such as BelleII, PANDA, the LHCb
upgrade, etc., hopefully guarantee continued excitement in
the near future.

\section*{Acknowledgements} We thank the PIC2012 committee
for organizing this wonderful conference and Prof. Roy A
Briere the convenor of our session for his help. This work
was supported in part by the National Natural Science
Foundation of China (NSFC) under Contract No. 11005115.

\end{document}